\documentclass[showpacs,11pt,english,pra,preprint,amsmath,amssymb]{revtex4}
\usepackage{graphicx}
\usepackage{bm}
\usepackage{graphicx}
\usepackage{epsfig}
\usepackage{graphics}
\usepackage{amsmath}
\date{\today}
\pacs{number}
\begin{document}
\title{Quantum Influence of Topological Defects in G\"odel-type Space-times}
\author{Josevi Carvalho}\email{josevi@ccta.ufcg.edu.br}
\affiliation{Unidade Acad\^emica de Tecnologia de Alimentos, Centro de Ci\^encias e Tecnologia Agroalimentar, 
Pereiros, 58840-000, Pombal, Para\'\i ba, Brazil.}

\author{Alexandre M. de M. Carvalho}\email{alexandre@fis.ufal.br}
\affiliation{Instituto de F\'{\i}sica, Universidade Federal de Alagoas, Campus A. C. Sim\~oes - Av. Lourival Melo Mota, s/n, Tabuleiro 
do Martins - Macei\'o - AL, CEP: 57072-970}

\author{Claudio Furtado}\email{furtado@fisica.ufpb.br}
\affiliation{Departamento de F\'{\i}sica, CCEN,  Universidade Federal 
da Para\'{\i}ba, Cidade Universit\'{a}ria, 58051-970 Jo\~ao Pessoa, PB,
Brazil}

\begin{abstract}
In this contribution, some solutions of the Klein-Gordon equation in the G\"odel-type metrics with an embedded cosmic string are considered. The quantum dynamics of a scalar particle in three spaces whose metric is described by different classes of G\"odel solution, with a cosmic string passing through the spaces, is found. The energy levels and eigenfunctions of the Klein-Gordon operator are obtained. We show that these eigenvalues and eigenfunctions depend on the parameter characterizing the presence of a cosmic string in the space-time. We note that the presence of topological defects breaks the degeneracy of energy levels.
\end{abstract}
\maketitle
\section{Introduction}
One of the most important solutions in the general relativity is the G\"odel metric~\cite{godel}, representing itself as a first cosmological solution with rotating matter. This solution is stationary, spatially homogeneous, possessing cylindrical symmetry, and its highly nontrivial property consists in breaking the causality implying in the possibility of the closed timelike curves (CTCs), whereas, as it was conjectured by Hawking~\cite{hawking}, the presence of CTCs is physically inconsistent. Furthermore, in~\cite{reboucas,reboucas2,reboucas3} this metric was generalized in cylindrical coordinates and the problem of causality was examined with more details, thus it turned out to be that one can distinguish three different classes of solutions. These solutions are characterized by the following possibilities: (i) there is no CTCs, (ii) there is an infinite 
sequence of alternating causal and non-causal regions, and (iii) there is only one non-causal region. In the paper~\cite{dabrowski} the quantities called super-energy and super-momentum which can be used as criteria of possibility of existence the CTCs were introduced. In~\cite{barrow} the CTC solutions in the G\"odel space are discussed within the string context~\cite{bertolami}. Another reasons for the interest to the G\"odel solution consist in the fact that the G\"odel universe allows for non-trivially embedded black holes~\cite{xe}. Different aspects of the G\"odel solutions are discussed also
in~\cite{barrow2}.

Recently, Drukker, Fiol and Simon~\cite{fiol} have investigated a close relation between a class of G\"odel solutions for general relativity in $(3+1)$-dimensions  and the Landau problem in a space of a constant curvature. They solved the Klein-Gordon equation in this curved background and observe the similarity of the energy levels with the Landau problem in flat, spherical and hyperbolic spaces. Guided by the analogy with the Landau problem, they have speculated on the possible
holographic description of a single chronologically safe region. In Ref.~\cite{gegenberg} the same relation between the flat G\"odel  solution (Som and Raychaudhuri limit) and Landau problem in the flat space was obtained. In other way the Landau problem has investigated in several situations, for example, in a hyperbolic space~\cite{comtet,dunne}, in a spherical space~\cite{dunne}, in the presence of topological defect~\cite{cflandau1,cblandau,jcmp}, for neutral 
particle~\cite{ericsson,linco1,linco2,knutpra} and for non-inertial frames in cosmic string background~\cite{prdknut}. 

The cosmic string ~\cite{Vil1,His} is one of the most important examples of the topological defects.
In general, they can arise in a gauge theory  with the spontaneous symmetry  breaking. They are predicted in some unified theories of particles interactions. Perhaps, they have been formed at phase transitions in the very early history of the Universe~\cite{Kib}. Other examples of such topological defects are the domain wall~\cite{Vil1} and the global monopole~\cite{Bar}. In particular, cosmic strings provide a bridge between the physical theories for microscopic and macroscopic scales. The richness of the new ideas they brought along to general relativity seems to justify the interest in the study of these structures and especially the role played by their topological features for the quantum systems.

In an astrophysical model, a topological defect can appear as an isolated object in an empty space. Among these defects, we consider the cosmic string. In this case the conical nature of the space-time outside the defect can produce interesting physical effects. One can generalize solutions of Einstein equations
in order to include the presence of cosmic string. We use a cut-gluing  method to build a solution like cosmic string, removing an angular sector $2\pi(1-\alpha)$. 
Imposing the requirement that the azimuthal angle about the axis of symmetry of the defect varies in the range $ 0 < \phi <2 \pi \alpha $, with $\alpha =1-4\lambda $ is the disclination parameter or the deficit angle of the conical geometry, and $\lambda$ is the linear mass density of the string. 
So, gluing the border radii of the disk with the wedge removed, we obtain the space-time with a cosmic string passing through it. The metric of a cosmic string in polar coordinates $(t,r,\phi,z)$ is given by 
\begin{eqnarray}
ds^{2}=-dt^{2} + dr^{2}+\alpha^{2}r^{2}d\phi^{2}+ dz^{2}.
\end{eqnarray}
Recently, several examples of studies in a space-time with a cosmic string were done revealing the role played by this topological defect in curved space-time. We can cite as examples, first,  Schwarzschild space-time with a cosmic string~\cite{mukunda,germano}, second, the Kerr space-time with a cosmic
string~\cite{22}, third, the  cosmic string in the background of AdS space~\cite{demellosahar}. In Ref. ~\cite{sandro}, 
a scalar quantum particle confined in two concentric thin shells in the Kerr-Newman, G\"odel and Friedmann-Robertson-Walker space-times with a
cosmic string passing through them has been considered~\cite{fiol}. In this contribution we discuss an example of a scalar quantum particle in a class of G\"odel space-time with 
a cosmic string passing trough it. We obtain the eigenvalues and eigenfunctions of energy. We demonstrate that the presence of cosmic string 
modifies the energy levels and breaks the degeneracy of eigenvalues. 

This paper is organized as follows. In the next section we present the various solutions of the Einstein fields equations 
related to the values of the parameter $l^2$ that can assume zero, positive and negative values. In these cases, the associated 
geometry is flat, hyperbolic and spherical, respectively. In the Section $III$, we analyse the solutions of the Klein-Gordon
equation for the G\"odel space-time for three possible values assumed by $l^2$. We also compare the results obtained
for the structures of the Landau levels (LL), discussing the degeneracy of these levels in the three cases. And in the 
concluding remarks, we presents a discussion of the main results we obtained.

\section{A G\"odel-type solution}
We find that G\"odel-type space-time, despite it is inadequate to describe our universe, have allowed to study many physical and mathematical models concerning 
gravitational backgrounds with rotation, violation of causality in general relativity, furthermore, the G\"odel-type {\bf metric has} the advantage that 
it is of rather compact form, thus, most of the calculations can be carried out analytically. A generalized discussion on the homogeneity and isotropy of the Riemannian manifolds was presented by Rebou\c cas and Tiomno~\cite{reboucas} with a G\"odel-type metric characterized by the parameters $(l,\Omega)$. These results lead to new solutions of the Einstein fields equations which can be called the G\"odel-type solutions.

Rebou\c{c}as and Tiomno built a solution for a metric G\"odel-type characterized by a vorticity, which represents a generalization of the original G\"odel metric. Since the symmetry imposed by the cosmic strings, it is natural to use the cylindrical coordinates to describe this space-time. Thus, the general G\"odel-type metrics with the presence of cosmic strings in polar coordinates $(t,r,\phi,z)$  can be written as
\begin{eqnarray}
\label{geoplain}
ds^2=-\left(dt+\alpha\Omega\frac{\sinh^2lr}{l^2}d\phi\right)^2+\alpha^2\frac{\sinh^22lr}{4l^2}d\phi^2+dr^2+dz^2.
\end{eqnarray}
The variables $(r, \phi, z, t)$ can take, respectively, the following values: {\bf $0\leq r <\infty$, $0\leq\phi\leq 2\pi$, $-\infty<(z,t)<\infty$.}
The parameter $\Omega$ characterizes the vorticity of the space. Note that the presence of topological defect does not change the conditions for existence
of closed timelike curves (CTCs) in the metrics (\ref{geoplain}). This condition is the same for G\"odel-type solution in absence of topological defect and 
is given by
\begin{eqnarray}
\tanh( lr_{c})=\frac{l}{\Omega}.
\end{eqnarray}
where $r_{c}$ is a critical radius outside of which the CTCs can exist. The region $r<r_{c}$ is denominated as the chronologically safe one. In the expression (\ref{geoplain}) we use natural units $c=G=1$. It is well known that in the asymptotic limit $l\rightarrow 0$ the metric has the same geometry as
the Som-Raychaudhuri space-time~\cite{som}. This solution of Einstein field equations recently appeared as G\"odel-type solution in string theory.
Another interesting situation is obtained when $\Omega=0$, in this case the system has a zero vorticity, and the disclination parameter is $\alpha=1$.
These both considerations make the metric~(\ref{geoplain}) to reduce to the Minkowski one. The original solution obtained by G\"odel~\cite{godel} can be
recovered for $l^2=\Omega^2/2$ and $\alpha=1$ and the anti-de Sitter case~\cite{rooman} corresponds to $l^2=\Omega^2$ and $\alpha=1$ case. 
As it was pointed by Drukker, Fiol and Simon~\cite{bfiol}, the physics of these kind of metrics is close related with the problem of a charged particle coupled to a magnetic field. The G\"odel-type metrics can be written in a general form,
\begin{eqnarray}
ds^2=-(dt+A_{i}(x)dx^i)^2+h_{ij}dx^idx^j.
\end{eqnarray}
where the spatial coordinates of the space-time are represented by the $x^i$. This general form represents surfaces of constant curvature in all cases  $l^2<0$, $l^2=0$ and $l^2>0$ as will be shown further. The main important features of this metrics is that the geodesics in these space-time 
are circles which have a physical description analogous to the orbits of electron moving in the orthogonal magnetic field known as the Larmor orbits. 
Furthermore, the energy levels in these space-times have a Landau structure as in the Minkowski space-time with a constant curvature~\cite{fiol}.  
In the next sections we will study the influence of the cosmic string parameters in this dynamics of scalar quantum particle.

\section{The Klein-Gordon Equation}
Now we will  investigate a scalar quantum particle in a G\"odel-type space-time. The relativistic quantum 
dynamics of a free spinless particle of mass $M$ is described by the Klein-Gordon equation. In its covariant form, this
equation takes the following form,
\begin{eqnarray}\label{kg}
\left[\frac{1}{\sqrt{-g}}\partial_{\mu}(\sqrt{-g}g^{\mu\nu}\partial_{\nu})-M^2\right]\Psi(t,\vec r)=0.
\end{eqnarray}
with $g$ being the determinant of metric tensor with inverse $g^{\mu\nu}$, and $\partial_{\mu,\nu}$ are derivatives with respect to spatial coordinates. The $\Psi(t,\vec r)$ is the amplitude of probability to find the particles around the $\vec r$ position at the time $t$.

In this contribution, we are interested in studying solutions of the Klein-Gordon equation for the background of 
G\"odel-type metric with a cosmic string embedded. We will study quantum effects produced by this geometry for eigenstates and eigenfunctions. A very important parameter that must be taken into account to obtain the solutions is the sign of $l^{2}$. The case $l^{2}=0$ corresponds to a flat solution, if only $\alpha=1$, otherwise corresponds to spinning cosmic string. On the other hand, the case $l^{2}<0$ corresponds to a spherically symmetric solution of positive curvature and finally, the case $l^{2}>0$ corresponds to a hyperbolic solution of negative curvature. These three cases will be discussed separately in the following sections.

\subsection{Scalar particle in Som-Raychaudhuri Space-time}
In this section we investigate the quantum dynamics of scalar quantum particle in a G\"odel space-time if $l^{2}= 0$ in 
(\ref{geoplain}), also well known by Som-Raychaudhuri solution~\cite{som}, so the metric (\ref{geoplain}) in this 
condition is reduced to
\begin{eqnarray}
\label{plain}
ds^2=-(dt+\alpha\Omega r^2d\phi)^2+\alpha^2r^2d\phi^2+dr^2+dz^2.
\end{eqnarray}
This solution has attracted much attention in string theory~\cite{horo,russo1,russo2} and in~\cite{boy,harm} it has been interpreted as a G\"odel-type solution in string theory. For this geometry, the Klein-Gordon equation (\ref{kg}) assumes the following form:
\begin{eqnarray}
\label{wave1}
\left[\frac{1}{r}\frac{\partial}{\partial{r}}\left(r\frac{\partial}{\partial{r}}\right)+\left(\frac{1}{\alpha r}\frac{\partial}{\partial{\phi}}-
\Omega r\frac{\partial}{\partial{t}}\right)^2-\frac{\partial^2}{\partial{t}^2}+\frac{\partial^2}{\partial{z}^2}-M^2\right]\Psi(r)=0,
\end{eqnarray}
This equation is independent of time and symmetrical by translations along the $z$-axis, as well by rotations, it is reasonable 
to write the solution as
\begin{eqnarray}
\label{ansatz}
\Psi(t,r,\phi,z) = e^{-iE t+im\phi+ikz}\Phi(r),
\end{eqnarray}
where $E$ and $m$ are constants of separation which can be interpreted as energy and angular momentum respectively. Substituting this {\it ansatz} into Eq.
(\ref{wave1}), we obtain the following differential equation for the radial function,
\begin{eqnarray}
\label{wave2}
\frac{d^2\Phi(r)}{dr^2}+\frac{1}{r}\frac{d\Phi(r)}{dr}-\left(\frac{m^2}{\alpha^2r^2}+\Omega^2E^{2}r^{2}\right)\Phi(r)+\left(E^2-\frac{2 m\Omega E}{\alpha}
-M^2-k^2\right)\Phi(r)=0.
\end{eqnarray}
We can also rewrite this equation in a more appropriate way: we introduce a new variable $\xi$, defined as $\xi=\Omega Er^2$, the resulting equation can be studied asymptotically, and its behaviour on critical points helps us to construct an equation in the same general form as the confluent hypergeometric equation. The result is
\begin{eqnarray}
\xi F''(\xi)+\left(\frac{|m|}{\alpha}+1-\xi\right)F'(\xi)-\left(\frac{|m|}{2\alpha}+\frac{1}{2}-\frac{\gamma}{4\Omega E}\right)F(\xi)=0.
\end{eqnarray}
with $\gamma=E^{2}-\frac{2m\Omega E}{\alpha}-k^2-M^2$. The solution of $F(\xi)$ is a polynomial of degree $n$ obtained by Frobenius method. Naturally, for all possible values of $\xi$ it is easy to observe that our solution is divergent at the extremal point. However this blow-up can be eliminated by a truncation on the series. This procedure is equivalent to imposing the following condition:
\begin{eqnarray}
 \left(\frac{|m|}{2\alpha}+\frac{1}{2}-\frac{\gamma}{4\Omega E}\right)=-n.
\end{eqnarray}
After some manipulations, we obtain the energy levels for the scalar particle in this background,
\begin{eqnarray}\label{enflat}
E = \left(2n+\frac{|m|}{\alpha}+\frac{m}{\alpha}+1\right)\Omega + \sqrt{\left(2n+\frac{|m|}{\alpha}+\frac{m}{\alpha}+
1\right)^2\Omega^2+M^2+k^2}.
\end{eqnarray}
For $\alpha=1$, the eigenvalues are reduced to the result found by Drukker, Fiol and Simon~\cite{bfiol,fiol}. One should notice that for $M=0$ and $k^{2}=0$ the energy levels (\ref{enflat}) are given by
\begin{eqnarray}\label{enflat1}
E=2\Omega\left(2n+\frac{|m|}{\alpha}+\frac{m}{\alpha}+1\right).
\end{eqnarray}
One should observe that the Eq.(\ref{enflat1}) the {\bf eigenvalues are similar} to those ones obtained for Landau levels in the presence of a cosmic string~\cite{cflandau1}. The degeneracy of energy levels (\ref{enflat}) is broken by the presence of the topological defect. After the normalization of the confluent hypergeometric function, we can write the expression for the eigenfunction as
\begin{equation}
\Psi(t,r,\phi,z)=C_{n,m}e^{-iEt+im\phi+ikz}r^{\frac{|m|}{2\alpha}}e^{-\frac{\Omega E}{2} r^2}
F\left(-n,\frac{|m|}{\alpha}+1,\Omega E r^2\right).
\end{equation}
where the $C_{n,m}$ represents the normalization constant.

Due to the presence of cosmic string, the topological parameter characterizes the eigenvalues as well as the 
eigenfunctions for the particle interacting with the conical geometry in G\"odel-type space-time. It is evident that, in order to obtain a complete description of the interaction between scalar quantum particle and gravitational fields produced by cosmological objects, we take into account not only the local features of the background, but also the topological features.
\subsection{Scalar Particle in Spherical Symmetrical G\"odel Space-time}\label{spher}

Now, we consider the limit of (\ref{geoplain}) where we can obtain a class of solutions of G\"odel-type possessing the spherical symmetry. We suggest that $l^2<0$ and introduce the new coordinates in (\ref{geoplain}): 
$R=\imath /2l$ and $\theta=r/R$. In the case when the sign of $l^2$ is negative, 
the metric (\ref{geoplain}) takes the following form,
\begin{eqnarray}
\label{mspher}
 ds^2=-\left(dt+\frac{\alpha\Omega r^2}{1+r^2/4R^2}d\phi\right)^2+\left(1+\frac{r^2}{4R^2}\right)^{-2}(dr^2+\alpha^2 r^2d\phi^2)+dz^2.
\end{eqnarray}
Note that the second term in (\ref{mspher}) corresponds to the metric of a two-sphere with a conical defect. In this background the Klein-Gordon 
equation can be written as
\begin{eqnarray}
&&\left[\left(1+\frac{r^2}{4r^2}\right)^2\frac{1}{r}\partial_r(r\partial_r)+\frac{\partial_\phi^2}{\alpha^2 r^2}+
\frac{1}{16R^4}\left(\frac{\partial_{\phi}}{\alpha}-4\Omega R^2\partial_{t}\right)^2r^2-\partial_t^2-\partial_z^2-
\frac{2\Omega \partial_\phi \partial_t}{\alpha}+\right.\nonumber\\
&+&\left.\frac{\partial_\phi^2}{2\alpha^2 R^2}\right]\Psi(t,\vec r)=0.
\end{eqnarray}
It is easy to see that this differential equation also has a translational symmetry along the z-axis and an azimuthal symmetry. This allows us once again to write the solution as
\begin{eqnarray}\label{anze}
\Psi(t,r,\phi,z)=e^{-iE t+im\phi+ikz}\Phi(r),
\end{eqnarray}
Following this procedure, we obtain a differential equation involving only the radial variable:
\begin{eqnarray}
\left(1+\frac{r^2}{4R^2}\right)^2\left[\frac{d^2}{dr^2}+\frac{1}{r}\frac{d}{dr}\right]\Phi(r)-\left[\frac{m^2}{\alpha^2 r^2}+\frac{1}{16R^4}\left(\frac{m}{\alpha}+4\Omega R^2E\right)^2r^2-\gamma'\right]\Phi(r)=0,   
\end{eqnarray}
where  $\gamma'=E^2-\frac{2\Omega mE}{\alpha}-k^2-M^2-\frac{m^2}{2\alpha^2R^2}$. By the stereographic representation of the sphere it is possible to introduce the following change of variables; $r=2R\tan\theta$. This transformation leads to the following differential equation,
\begin{eqnarray}
\Phi''(\theta)+\left(\frac{1}{\sin\theta \cos\theta}-\frac{2\sin\theta}{\cos\theta}\right)\Phi'(\theta)-\left(\frac{a^2\cos^2\theta}{\sin^2\theta}+
\frac{b^2\sin^2\theta}{\cos^2\theta}-4R^2\gamma'\right)\Phi(\theta)=0,
\end{eqnarray}
where $a=\alpha/m$ and $b=(m/\alpha+4\Omega R^{2}E)$. By making two changes of variables consequently, with the first one is
$x=\cos\theta$, and the second one is $\xi=1-x^{2}$, we rewrite this equation as
\begin{eqnarray}
\xi(1-\xi)\frac{d^2\Phi}{d\xi^2}+(1-2\xi)\frac{d\Phi}{d\xi}&-&\left[\frac{m^2(1-\xi)}{4\alpha^2 \xi}+\frac{\lambda^2\xi}{4(1-\xi)}
-R^2\gamma'\right]\Phi=0,
\end{eqnarray}
where $ \lambda=\left|\frac{m}{\alpha}+4\Omega R^2E\right|$. Let us require again that the solution must be finite at 
$\xi=0$ and $\xi=1$. We can write the solution as $\Phi(\xi)=(1-\xi)^\gamma\xi^\beta F(\xi)$,
with the parameters $\beta=\left|\frac{m}{2\alpha}\right|,\gamma=\frac{\lambda}{2}$. This new transformation leads to the following differential equation for the function $F(\xi)$,
\begin{eqnarray}
\label{hip1}
&&\xi(1-\xi)\frac{d^2 F}{d\xi^2}+\left[\frac{|m|}{\alpha}+1-\left(\frac{2|m|}{\alpha}+4\Omega R^2E+2\right)\xi\right]\frac{dF}{d\xi}-\left[\frac{m^2}{\alpha^2}+2\Omega R^2\left(\frac{|m|}{\alpha}+\frac{m}{\alpha}+1\right)E\right. \nonumber 
\\ &+& \left. \frac{|m|}{\alpha}-R^2(k^2+M^2)\right]F(\xi)=0.
\end{eqnarray}
We can identify that the above equation coincides with the general form of the hypergeometric equation. Making an identification with a general hypergeometric function $F(A,B,C,\xi)$, satisfying the equation $x(1-x)F''(x)+(C-(A+B+1)x)F'(x)-ABF(x)=0$, we find two first parameters $(A,B)$ of the hypergeometric series to be
\begin{eqnarray}
(A,B)&=&\frac{1}{2}\left(1+\frac{2|m|}{\alpha}+4\Omega R^2E\right)\pm
\nonumber\\&\pm&
\frac{1}{2}\sqrt{1+4\Omega R^2E+2\Omega R^2\left(\frac{|m|}{\alpha}-\frac{m}{\alpha}\right)E+4R^2(E^2-k^2-M^2)}.
\end{eqnarray}
These parameters allow us to calculate the eigenvalues and eigenfunctions associated with the problem. This procedure is similar to that performed above. This enables us, after some manipulations, to find that energy levels are given by
\begin{eqnarray}
\label{spectra}
E_{n,m}&=&\left(2n+\frac{|m|}{\alpha}+\frac{m}{\alpha}+1\right)\Omega +\nonumber\\
&+& {\sqrt{\left(2n+\frac{|m|}{\alpha}+\frac{m}{\alpha}+1\right)^2\Omega^2+
\frac{1}{R^2}\left(n+\frac{|m|}{\alpha}\right)\left(n+\frac{|m|}{\alpha}+1\right)+M^2+k^2}}.\;\;\;\;\;
\end{eqnarray}
Degeneracy of $E$ is finite for the interval $-n\leq m/\alpha\leq 4\Omega R^2w$. Note  that the presence of the topological defect
reduces the degeneracy of energy levels since parameter $\alpha$ belongs to the range $0<\alpha <1$.

We see that again that the energy levels depend on the topological parameter $\alpha$. Therefore, all information on a background space-time can be obtained by studying the eigenvalues problem. It is worth to call the attention to the limit of Minkowskian background for the spherical case studied in this section. Note that in the limit $(R\to \infty)$, the eigenvalues (\ref{spectra})  reproduce the spectra of the particle 
in the Som-Raychaudhuri geometry with the presence of topological defect, and in the limit $\alpha=1$ we obtain the  eigenvalues of the scalar quantum particle in the Som-Raychaudhuri background~\cite{fiol,gegenberg,damiao}.

The corresponding eigenfunction is
\begin{eqnarray}
\Psi(t,r, \phi, z)=C_{n,m}e^{-iEt+im\phi+ikz}r^{\frac{|m|}{2\alpha}}(1-\Omega Er^2)^{\frac{1}{2}\left(\frac{|m|}{\alpha}+
4\Omega R^2E\right)}F\left(A,B,\frac{|m|}{\alpha}+1,\Omega E r^2\right).
\end{eqnarray}
The constants $C_{n,m}$ are obtained from the conditions of normalization of the wave function.

The degenerate states are present when we suppose the topological parameter to be integer. For non-integer values of $\alpha$ the degeneracy 
decreases in comparison with the case $\alpha=1$. Moreover, as in the Landau problem {\bf on surface} with a constant positive curvature, 
the spectra (\ref{spectra}) involve a linear term accompanied by the quadratic correction introduced by the curvature, which is in accordance with 
the results presented in~\cite{comtet, dunne} where the particles interacting with uniform 
magnetic field in the surface were considered.

\subsection{Hyperbolic Coordinates}\label{hyperb}
Now we investigate the hyperbolic solution of (\ref{geoplain}) where $(l^2>0)$, which implies a hyperbolic space-time given by
\begin{eqnarray}\label{mhyper}
ds^2=-\left(dt+\frac{\alpha\Omega r^2}{1-l^2r^2}d\phi\right)^2+(1-l^2r^2)^{-2}(dr^2+\alpha^2 r^2d\phi^2)+dz^2.
\end{eqnarray}
In this case, the radial equation obtained from the Klein-Gordon equation (\ref{kg}) describing a scalar particle in this space-time is written as follows:
\begin{eqnarray}
\label{kg3}
 \left[(l^2r^2-1)^2\left(\frac{d^2}{dr^2}+\frac{1}{r}\frac{d}{dr}\right)-\frac{m^2}{\alpha^2 r^2}-l^4\left(\frac{m}{\alpha}-
 \frac{\Omega E}{l^2}\right)^2r^2+\gamma'\right]\Phi(r)=0,
\end{eqnarray}
where $\gamma'=E^2+\frac{2l^2m^2}{\alpha^2}-\frac{2\Omega mE}{\alpha}-k^2-M^2$ and we have write $\Psi$ in the same form of Eq. (\ref{anze}). Again, one must perform a set of transformations of variables 
to transform the differential equation to a more suitable way. The first transformation involves the radial variable which
is now written as $r=\tanh(l\theta)/l$. It results in a new equation for the variable $\theta$,
\begin{eqnarray}
\left\{\frac{d^2}{d\theta^2}+\left[\frac{2l\sinh(l\theta)}{\cosh(l\theta)}+\frac{l}{\sinh(l\theta)\cosh(l\theta)}\right]\frac{d}{d\theta}-
\left[\frac{{\cal A}^2l^2\cosh^2(l\theta)}{\sinh^2(l\theta)}+\frac{l^2{\cal B}^2\sinh^2(l\theta)}{\cosh(l\theta)}-\gamma'\right]\right\}\Phi(\theta)=0,\;\;\;\;
\end{eqnarray}
where ${\cal{A}}=\frac{m}{\alpha}$ and ${\cal{B}}=\frac{\Omega E}{l^2}-\frac{m}{\alpha}$. We also must carry out the following sequence of changes 
of variables: first, we introduce $y=\cosh(l\theta)$, then we denote $\xi=y^{2}-1$. With these transformations, after some algebra we arrive at
\begin{eqnarray}\label{30}
\Phi''(\xi)+\frac{1+2\xi}{\xi(1+\xi)}\Phi'(\xi)-\left[\frac{\beta^2}{\xi^2}+\frac{\gamma^2}{(1+\xi)^2}-\frac{\gamma'}{4l^2\xi(1+\xi)}\right]\Phi(\xi)=0, 
\end{eqnarray}
which can be rewritten as
\begin{eqnarray}\label{31}
\xi(1+\xi)\Phi''(\xi)+(1+2\xi)\Phi'(\xi)-\left[\frac{\beta^2 \xi(1+\xi)}{\xi}+\frac{\gamma^2 \xi}{(1+\xi)}-\frac{\lambda^{2} -1}{4}\right]\Phi(\xi)=0, 
\end{eqnarray}
where $\frac{\gamma'}{4l^2}=\frac{\lambda^{2}-1}{4}$. The solution of this differential equation can be obtained for the ansatz,
\begin{eqnarray}
\label{phi3}
\Phi(r)=\xi^{\beta}(1+\xi)^{\gamma}F(\xi),
\end{eqnarray}
where $\beta=|m|/{2\alpha}$ and $\gamma=(|m|/\alpha-\Omega E/l^2)/2$. The solution (\ref{phi3}) is consistent with the asymptotic conditions discussed above. Replacing (\ref{phi3}) into (\ref{31}), we obtain
\begin{eqnarray}
\label{hip2}
\xi(1-\xi)\frac{d^2F}{d\xi^2}+\left[\frac{|m|}{\alpha}+1-\left(\frac{2|m|}{\alpha}-\frac{\Omega E}{l^2}+2\right)\xi\right]\frac{dF}{d\xi}
&-&\left[\frac{m^2}{\alpha^2}-\frac{\Omega}{2l^2}\left(\frac{|m|}{\alpha}+\frac{m}{\alpha}+1\right)+\frac{|m|}{\alpha}-\frac{\Omega E}{2l^2}\right.
\nonumber \\ &+& \left. \frac{E^2-M^2-k^2}{4l^2}\right]F=0.
\end{eqnarray}
This differential equation can be easily identified with the hypergeometric differential equation. The parameters characterizing the hypergeometric function $F(a, b, c, x)$ are given by
\begin{eqnarray}
(a,b)=\frac{1}{2}\left(1+\frac{2|m|}{\alpha}-\frac{\Omega E}{l^2}\right)\pm\frac{1}{2}\sqrt{1+\frac{\Omega^2E^2}{l^4}-
\frac{\Omega}{2l^2}\left(\frac{|m|}{\alpha}-\frac{m}{\alpha}\right)E-\frac{E^2-k^2-M^2}{l^2}}.         
\end{eqnarray}
The truncation of the series allows us to obtain energy spectrum
\begin{eqnarray}
\label{freq}
E_{n,m}&=&\left(2n+\frac{|m|}{\alpha}+\frac{m}{\alpha}+1\right)\Omega  + \\ \nonumber &+&\sqrt{\left(2n+\frac{|m|}{\alpha}+\frac{m}{\alpha}+1\right)^2\Omega^2
-4l^2\left(n+\frac{|m|}{\alpha}\right)\left(n+\frac{|m|}{\alpha}+1\right)+M^2+k^2}.
\end{eqnarray}
In the hyperbolic geometry the energy spectrum can assume discrete or continuous values~\cite{comtet,damiao}. The existence of these two cases is associated with the condition which parameter $\lambda$ must obey~\cite{fiol,damiao}. This parameter is defined by
\begin{eqnarray}
\lambda^2=1+\frac{\Omega^2-l^2}{l^4}E^2-\frac{\Omega}{2l^2}\left(\frac{|m|}{\alpha}-\frac{m}{\alpha}\right)E+\frac{k^2+M^2}{l^2}
\end{eqnarray} 
It may assume values either $\lambda>1$ or $\lambda<1$. The condition $\lambda>1$ is satisfied when $\Omega^2>l^2$, and, as it was discussed by~\cite{damiao},
{\bf the corresponding spectra} belongs to a region with discrete energy levels. For $\Omega^2<l^2$, the discrete energy {\bf levels are} bounded from above by the energy
\begin{eqnarray}\label{upper}
E\leq\frac{\Omega l^2}{l^2-\Omega^2}\left(\frac{|m|}{\alpha}-\frac{m}{\alpha}\right)\pm\sqrt{\frac{\Omega^2l^4}{(l^2-\Omega^2)^2}\left(\frac{|m|}{\alpha}
-\frac{m}{\alpha}\right)^2+\frac{l^2(k^2+m^2)}{l^2-\Omega^2}}. 
\end{eqnarray}
Note that, the  condition (\ref{upper}) is satisfied for upper sign case by positive energy and for the case of lower sign is satisfied by negative energy.
{\bf Above} this limit, the spectrum is continuous. 

The eigenstates corresponding to the discrete spectra can be written in the form
\begin{eqnarray}
 \Psi(t,r,\phi,z) = C_{n,m}e^{-iEt+im\phi+ikz}r^{\frac{|m|}{2\alpha}}(1+\Omega E r^2)^{\frac{1}{2}\left(\frac{|m|}{2\alpha}-\frac{\Omega E}{l^2}\right)}
 F\left(a, b,\frac{|m|}{\alpha}+1, \Omega E r^2\right),
\end{eqnarray}
for $0\leq m/\alpha\leq\infty$, and
\begin{eqnarray}
 \Psi(t,r,\phi,z) = C_{n,m}e^{-iEt+im\phi+ikz}r^{-\frac{|m|}{2\alpha}}(1+\Omega E r^2)^{-\frac{1}{2}\left(\frac{|m|}{2\alpha}- '\frac{\Omega E}{l^2}\right)}F\left(a, b, \frac{|m|}{\alpha}+1, \Omega E r^2\right),
\end{eqnarray}
for $-n\leq m/\alpha\leq0$.

{\it Otherwise, we have {\bf continuous spectra}, where the condition for $n$ following from the restriction $\lambda<1$ is given by the expression}
\begin{eqnarray}
\label{lim}
\left(2n+\frac{|m|}{\alpha}+\frac{m}{\alpha}+1\right)\leq\frac{\Omega}{l}\sqrt{\frac{M^2+k^2}{l^2-\Omega^2}-
\frac{\Omega}{2}\left(\frac{|m|}{\alpha}-\frac{m}{\alpha}\right)},
\end{eqnarray}
and above this limit the continuous {\bf solutions appear} with the energy spectra obeying the equation (\ref{freq}).

In order to summarize this section, we call the attention to the fact that we obtained two distinct behaviours of the energy spectra. One case corresponds to the region described by $\Omega^2>l^2$, which the closed timelike curves can present in a classical background. The spectrum of the energy associated with this region is discrete. The other observed behaviour corresponds to $\Omega^2<l^2$. The corresponding space-time is free of closed time-like curves. In this case we have a discrete energy levels upper-bound by a continuous spectrum obeying Eq.(\ref{upper}) and the quantum number $n$ is restricted by the condition (\ref{lim}).

\section{Conclusion}\label{conc} 

We studied the impacts produced by the presence of topological defect in the family of G\"odel-type space-times for the energy levels of a scalar quantum particle in this geometry. We have solved the Klein-Gordon equation in the spherical, hyperbolic and flat geometries and found that the eigenvalues and eigenfunctions depend of the parameter characterizing the presence of a cosmic string in the G\"odel-type background space-time. The presence of parameter $\alpha$ {\bf breaks} the degeneracy of energy levels in the three cases: Som-Raychaudhuri, 
spherical G\"odel and Hyperbolic G\"odel solutions.

Comparing our results with the conclusions of the paper~\cite{fiol}, we find that at the classical level all solutions corresponding to discrete energy with $n$ bounded from above by (\ref{lim}) for $\Omega^2 > l^2$ and unbounded in the opposite case, coincide with the results for the range of parameters $(\Omega, l)$ in which the closed time-like curves can exist for classical geodesics motion in the hyperbolic geometry, while the moving in regions free of CTCs with unbounded orbits is equivalent, at the quantum level,
to regions with continuous energy states. Even or this relation for classical and quantum levels we have another problem which has a physical content analogous to the G\"odel-type solution on hyperbolic geometries, that is, the Landau problem on surfaces of constant curvature. In this case the discrete energy levels are obtained for an enough strong magnetic field which forces the electron to follow a closed orbit. The unbounded orbits run by the particle are associated  with the continuous energy spectra, and the classical motion is free of closed time-like 
curve as it was discussed earlier. We finalize this contribution calling the attention that the results with G\"odel-type metrics found in~\cite{bfiol, som} can be obtained in the limits $(R\to \infty, l^2 \to 0)$ and $\alpha =1$.

{\bf Acknowledgements.} \quad We thank CNPQ, CAPES, FAPESQ, CAPES, for the financial support.

\end{document}